\documentclass[journal]{IEEEtran}
\usepackage{graphicx}
\usepackage{color}
\usepackage{fancyhdr}
\usepackage[cmex10]{amsmath}
\usepackage{mdwmath}
\usepackage{amssymb}
\usepackage{textcomp}
\usepackage{times}
\usepackage{multicol}
\usepackage{float}
\usepackage[numbers,sort&compress]{natbib}
\usepackage{color, soul}

\usepackage{color}
\usepackage{algorithm} 
\usepackage{algorithmic} 
\usepackage{multirow}

\ifCLASSINFOpdf
\else
\fi
\begin{document}
%
\title{Outage Probability of Energy Harvesting Relay-aided Cooperative Networks Over Rayleigh Fading Channel}
%
%


\author{\authorblockN{Tao Li\authorrefmark{1}, Pingyi Fan, Senior Member, IEEE \authorrefmark{1} and Khaled Ben Letaief, Fellow, IEEE\authorrefmark{2}\\
\authorblockA{\authorrefmark{1}Tsinghua National Laboratory for Information Science and Technology, and \\
Department of Electronic Engineering, Tsinghua University, Beijing, China.\\
\authorrefmark{2}Department of Electrical and Computer Engineering, HKUST, Hong Kong.}\\
E-mails: \authorrefmark{1}\{litao12@mails, fpy@\}.tsinghua.edu.cn, \authorrefmark{2}eekhaled@ust.hk\\ } }

\maketitle
\thispagestyle{empty}

\begin{abstract}
Energy harvesting technique is a potential way for relay node energy supply in cooperative networks in terms of deployment flexibility and maintain charge reduction. Unlike traditional power source, relay node in this case may run out of energy with certain probability, which can degrade the benefit from relay-aided cooperative transmission. In this paper, we concentrate on the outage behavior of cooperative networks aided by energy harvesting relay node in slow fading channel, and attempt to derive the closed-form expression of outage probability of proposed cooperative protocol. Compared with traditional direct transmission protocol, two conclusions are derived: 1) the diversity gain cannot be increased excepting the extreme case that energy-exhausted probability is zero; 2) a multiplicative gain for improving system performance can be obtained in terms of minimizing outage probability.
\end{abstract}


%
\IEEEpeerreviewmaketitle

\section{Introduction}
%
%
%
%

\IEEEPARstart{I}{n} wireless communication scenario, fading is a great challenge for reliable information transmission \cite{Tse_1}. Especially for realtime applications in slow fading environment, where coherent time is on the order of transmission delay constraint requirement, outage event may occur if instantaneous signal to noise ratio (SNR) is below the minimal acceptable threshold value. Many advanced techniques have been proposed to alleviate this problem in the literature. Among them, relay-aided cooperative transmission is an efficient way, see e.g. \cite{Laneman_2,Lei_3,Xiong_4}. However, the flexibility of relay node deployment is limited by existing power supply. Besides, maintain costs, such as electric charge, also restricts the massive deployment.

Recently, energy harvesting, which can collect energy from renewable resources in ambient environment such as solar energy, wind, geothermal energy and even RF signal, has attracted much attention. It is considered as a promising approach for powering relay node, see e.g. \cite{Nasir_6,Huang_7,Ding_8,Chalise_9}. Specifically, the work in \cite{Nasir_6} discussed the optimal relaying protocols under two different receiver structures. Power allocation strategies in cooperative networks powered by harvested energy were considered in \cite{Huang_7,Ding_8}. The performance boundary of AF-MIMO system with energy harvesting receiver was analyzed in \cite{Chalise_9}. According to current state of art, sufficient energy can be provided in this way \cite{Gorlatova_10}. The only difference is that relay node in this case may run out of energy with certain probability due to the fluctuation of harvested energy, which can degrade the benefit from cooperative technology.
To the best of our knowledge, compared with relay-aided cooperative networks powered by stable power supply, the performance loss resulted from harvested energy fluctuation hasn't been considered in the literature. Moreover, whether it is profitable to employ energy harvesting relay compared with simple direct link transmission is even not given explicitly.

As a consequence, this paper attempts to evaluate the performance of energy harvesting relay-aided cooperative network in terms of outage probability. An on-off model is employed to characterize the energy flow harvested from surrounding environment, which can capture the stochastic property of it very well. The explicit closed-form expression of outage probability in the proposed cooperative protocol is obtained. By theoretical analysis, it indicates that energy harvesting relay node cannot increase the diversity gain of system excepting energy-exhausted probability is zero. But, though more time slots are needed in relay cooperative protocol compared with direct transmission protocol, the outage performance can be improved significantly in energy harvesting relay-aided protocol during the concerned SNR range, especially when energy-exhausted probability is small. Since energy-exhausted probability can be decreased to very small level in well-designed system \cite{Omur_11}, it can be concluded that it is feasible and profitable to employ relay node powered by harvested energy to improve system transmission performance.

\section{Preliminary}

\subsection{System Model}

\begin{figure}[!t]
\centering
\includegraphics[width=3.0 in]{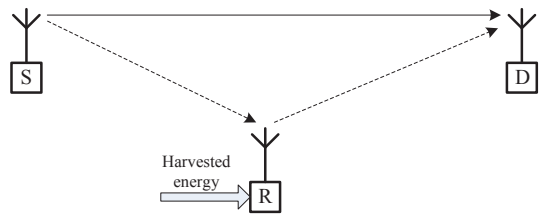}
\caption{A simple example of system structure for energy harvesting relay-aided cooperative network, in which solid line denotes direct link and dashed line denotes relay link.}
\label{fig:system framework}
\end{figure}

Fig. \ref{fig:system framework} illustrates an example of system structure for energy harvesting relay-aided cooperative network, which consists of one source node, one relay node and one destination node, denoted as $S$, $R$ and $D$, respectively. It is assumed that $S$ is powered by a traditional stabilized power source with constant power flow $P_s$. In contrast, $R$ is powered by the energy harvesting module with average power output flow $P_{av}$. Information flow is intended to be transmitted from $S$ to $D$. In direct transmission protocol, only the direct transmission link between $S$ and $D$ is available while both direct link and relay link are available in cooperative transmission protocol.

Assuming time and frequency synchronization is ideal, and channel side information are known prior by each terminals. $x_s(t)$ denotes the transmit signal at $S$ with zero mean and $\mathbb{E}[x_s(t)]=1$, where $\mathbb{E}[.]$ denotes statistical mean operation. For direct transmission link, the baseband-equivalent model for this channel is
\begin{equation}\label{eqn:channel model for direct transmission}
  y_{sd}(t)=h_{sd}\sqrt{P_s} x_{s}(t)+n_{d}(t)
\end{equation}
where $y_{sd}(t)$ denotes the received signal from direct link at $D$. The corresponding channel fading coefficient and additive noise are denoted as $h_{sd}$ and $n_{d}(t)$, respectively.
Besides, it is worth noting that channel coherent time is on the order of delay constraint and $h_{sd}$ doesn't change during a whole signal block in slow fading channel.

For relay transmission link with enough energy supply at $R$, in order to avoid interference, the whole transmission phase should be divided into two sub-phases. In the first half phase, information is transmitted from $S$ to $R$. Similar to the result in (\ref{eqn:channel model for direct transmission}), it can be expressed as
\begin{equation}\label{eqn:channel model for relay transmission 1}
   y_{sr}(t)=h_{sr}\sqrt{P_s} x_{s}(t)+n_{r}(t)
\end{equation}
where $y_{sr}(t)$ is received signal and $n_{r}(t)$ is additive noise at $R$.

Then, the received information is retransmitted from $R$ to $D$ with constant transmit power $P_r$ in the second half phase, which is expressed as
\begin{equation}\label{eqn:channel model for relay transmission 2}
   y_{rd}(t)=h_{rd}\sqrt{P_r} x_{r}(t)+n_{d}(t)
\end{equation}
where $x_{r}(t)$ denotes the transmit signal at $R$.

In (\ref{eqn:channel model for direct transmission}-\ref{eqn:channel model for relay transmission 2}), $h_{ij}$ captures the effect of path loss, shadowing and small-scale fading, where $i\in \{s,r\}$ and $j\in\{r,d\}$. Since small-scale fading is the main factor for outage event, this paper concentrates on the effect of small-scale fading upon transmission performance in terms of outage probability. Without loss of generality, $h_{ij}$ is modeled as a zero-mean, independent, circularly symmetric complex Gaussian random variables with unit variance. Besides, $n_j(t)$ is also modeled as a zero-mean, independent, circularly symmetric complex Gaussian random variables with variance $\sigma_0^2$.

Based on above assumptions, if direct transmission protocol is employed, only direct link $S-D$ is available. The overall data rate from $S$ to $D$ is given in the following, which can be achieved by zero-mean, circularly symmetric complex Gaussian inputs.
\begin{equation}\label{eqn:instantaneous information rate in direct transmission}
  R_{sd}=W \log_2 (1+\frac{|h_{sd}|^2 P_s}{\sigma_0^2})
\end{equation}
where $W$ denotes the frequency bandwidth.

Assuming the minimum acceptable rate is $R_0$, the outage probability $Pr[R_{sd}<R_0]$ in this case can be expressed as
\begin{equation}\label{eqn:outage probability in direct transmission}
  p_{sd}^{out}=1-exp(-\frac{2^{R_0/W}-1}{P_s / \sigma_0^2})
              \approx \frac{2^{R_0/W}-1}{P_s / \sigma_0^2}
\end{equation}
where the last step is tight when $P_s / \sigma_0^2$ is large enough.

\subsection{Energy Harvesting Modeling}

Due to fluctuation of renewable resource, the value of instantaneous harvested energy is always time-varying. So it cannot maintain a stable energy output flow just like conventional power supply. It is modeled as a stationary stochastic variable in \cite{Omur_11} while a deterministic model is used to describe the energy flow when the energy arrival time and the harvested amount are known prior to transmitter in \cite{Huang_7}.

In this paper, it is assumed $R$ transmits signal with constant power $P_r$. Thus, it is realistic to utilize an on-off model to characterize the harvested energy flow from a statistical perspective. Namely, if the available energy at relay node $R$ is more than $P_r T_0$ at the beginning of each signal block, where $T_0$ denotes the length of a whole signal block, signal can be transmitted by $R$ successfully. Otherwise, signal cannot be transmitted from $R$, the probability of which is called as energy-exhausted probability and denoted as $p_{ex}$.

Thus, energy harvesting modules can be characterized by the pair $(P_r,p_{ex})$, which can captures the stochastic property of harvested energy. Energy storage at $R$, such as battery, can decease the value of $p_{ex}$. In particular, energy-exhausted probability can be asymptotic to zero with respect to average power if energy storage capacity is big enough \cite{Omur_11}. Besides, it is worth noting that specific probability distribution of harvested energy depends on the specific sort of renewable source employed, which is beyond the scope of this paper.

\section{Cooperative Protocols and Outage Behavior Analysis}

\subsection{Cooperative Transmission Protocol}

Since the exact capacity region of relay channel is still an open problem \cite{Laneman_2} and the main purpose of this paper is to evaluate the benefit generated by energy harvesting relay-aided cooperative transmission, we only consider some simple relay strategy instead of attempting to obtain some optimal transmission strategies. Specifically, when $h_{sd}^2$ is bigger than the threshold value shown in (\ref{eqn:outage probability in direct transmission}), only direct link is used to transmit signal, which is just the same as that in direct transmission protocol. However, when $h_{sd}^2$ is smaller than the threshold value, relay link should activated to provide an alternative transmission link instead of direct link. From a view of practical system, this strategy is feasible since relay link should be used as less as possible to decrease the overhead information expenditure.

Due to fluctuation of energy harvesting, relay link cannot be always started up successfully, the failure probability of which is $p_{ex}$. On the condition that available energy at $R$ is enough for signal transmission, $R_{re}$ denotes the data rate that relay link $S-R-D$ can support, and $p_{re}^{out}$ denotes the outage probability $Pr[R_{re}<R_0]$ for relay link. According to above assumptions, the overall outage probability under the proposed cooperative protocol can be expressed as
\begin{equation}\label{eqn:outage probability in cooperative transmission}
  p_{co}^{out}=p_{sd}^{out}\cdot [p_{ex}+(1-p_{ex})\cdot  p_{re}^{out}]
\end{equation}

Let's attempt to derive the expression of $p_{re}^{out}$. It is assumed that amplify-and-forward relay scheme is employed at relay node. The relationship between received signal $y_{sr}(t)$ and retransmitted signal $x_r(t)$ at $R$ in this case can be expressed as
\begin{equation}\label{eqn:constraint at relay node}
  x_r(t)=\beta y_{sr}(t)
\end{equation}
where $\beta$ is power normalized factor \cite{Laneman_2}
\begin{equation}\label{eqn:amplify factir at relay node}
  \beta=\sqrt{\tfrac{1}{|h_{sr}|^2 P_s+\sigma_0^2}}
\end{equation}

Combining the results in (\ref{eqn:channel model for relay transmission 1}-\ref{eqn:channel model for relay transmission 2}) and (\ref{eqn:constraint at relay node}-\ref{eqn:amplify factir at relay node}), the instantaneous information rate over relay link $S-R-D$ in this case is
\begin{equation}\label{eqn:instantaneous information rate in relay transmission}
  R_{re}=\frac{1}{2} W \log_2 (1+f(h_{sr},h_{rd}))
\end{equation}
where the multiplicative factor $\tfrac{1}{2}$ is resulted from channel resource division between the link $S-R$ and $R-D$, and the expression of $f(h_{sr},h_{rd})$ is
\begin{equation}\label{eqn:instantaneous snr}
  f(h_{sr},h_{rd})=\frac{\frac{|h_{sr}|^2 P_s}{\sigma_0^2} \cdot \frac{|h_{rd}|^2 P_r}{\sigma_0^2}}{\frac{|h_{sr}|^2 P_s}{\sigma_0^2}+\frac{|h_{rd}|^2 P_r}{\sigma_0^2}+1}
\end{equation}

\subsection{Outage performance analysis}

Based on the literature \cite{Xiong_4,Ding_5}, the cumulative distribution function (CDF) of $f(h_{sr},h_{rd})$ has been given, which is concluded as following Lemma (see detailed proof in the Appendix of \cite{Ding_5}).

\emph{Lemma 1: Provided that both $h_{sr}$ and $h_{rd}$ are  zero-mean, independent, circularly symmetric complex Gaussian random variables with unit variance, the CDF of $f(h_{sr},h_{rd})$ in (\ref{eqn:instantaneous snr}) can be shown as}
\begin{equation}\label{eqn:the CDF of function F}
\begin{split}
  P[f<z]= & 1-e^{-(\frac{\sigma_0^2}{P_s}+\frac{\sigma_0^2}{P_r})z} \sqrt{4\tfrac{\sigma_0^4}{P_s P_r} z(z+1)} \\
                         & \,\,\, \cdot K_1(\sqrt{4\tfrac{\sigma_0^4}{P_s P_r} z(z+1)})
\end{split}
\end{equation}
\emph{where $K_1(x)$ is the first order modified Bessel function of the second kind.}

In order to obtain the closed-form of outage probability and make the comparison convenient, some approximation relationship in high SNR region as shown in (12) can be employed to simplify the expression.
\begin{equation}\label{eqn:approximate relationship}
  \lim \limits_{x\rightarrow0} K_1(x)=\tfrac{1}{x} \,\,\, \text{and} \,\,\, \lim \limits_{x\rightarrow0} 1-e^{-x}=x
\end{equation}

Substituting the results in (\ref{eqn:instantaneous information rate in relay transmission}-\ref{eqn:approximate relationship}) into (\ref{eqn:outage probability in cooperative transmission}), the explicit closed-form expression of outage probability for cooperative transmission protocol in high SNR range powered by harvested energy can be expressed as
\begin{equation}\label{eqn:overall outage probability in cooperative transmission}
  p_{co}^{out}=\frac{\sigma_0^2}{P_s}(2^{R_0/W}-1) \cdot \{p_{ex}+(\frac{\sigma_0^2}{P_s}+\frac{\sigma_0^2}{P_r})(2^{2R_0/W}-1)(1-p_{ex})\}
\end{equation}

Since diversity gain is an important metric for the performance of cooperative networks, let's compare the performance of three different systems in terms of diversity gain, namely direct transmission system, relay-aided cooperative system powered by stable supply and cooperative system powered by energy harvesting.
For making below description more clear, it is assumed that $P_s=P_r=P_0$ in the sequel. Under this condition, the expression of outage probability in (\ref{eqn:overall outage probability in cooperative transmission}) can be rewritten as
\begin{equation}\label{eqn:outage probability evaluation}
\begin{split}
  p_{co}^{out}=& (2^{R_0/W}-1)\cdot p_{ex} \cdot \rho^{-1}+2(2^{R_0/W}-1)\cdot \\
               & (2^{2R_0/W}-1)\cdot (1-p_{ex})\rho^{-2}
\end{split}
\end{equation}
where $\rho$ denotes instantaneous SNR at each receiver
\begin{equation}\label{eqn:the expression of snr}
  \rho=\tfrac{P_0}{\sigma_0^2}
\end{equation}

Recalling the definition of diversity gain from \cite{Ding_5} as in (\ref{eqn:definition of diversity gain}), some conclusions can be derived based on the results in (\ref{eqn:outage probability evaluation}), which is introduced as following corollaries.
\begin{equation}\label{eqn:definition of diversity gain}
  d = -\lim \limits_{\rho\rightarrow\infty} \frac{\log (p_{co}^{out})}{\log \rho}
\end{equation}

\emph{Corollary 1: The diversity gain for proposed relay-aided cooperative networks can be expressed as}
\begin{equation}\label{eqn:diversity gain of cooperative system}
  d=
\begin{cases}
  2, & \textrm{if $p_{ex}=0$}  \\
  1, & \textrm{otherwise}
\end{cases}
\end{equation}
\begin{proof}
Substituting (\ref{eqn:outage probability evaluation}) into (\ref{eqn:definition of diversity gain}) can get the result.
\end{proof}

It can be concluded that diversity gain can not be increased comparing with direct transmission protocol excepting $p_{ex}=0$. Specifically, when $p_{ex}=0$, it is equivalent to the cooperative system powered by stable supply $P_r$, and the corresponding diversity gain is $2$, which is consistent with the results proposed in \cite{Laneman_2}. Otherwise, diversity gain is just $1$. Besides, it can be proved that diversity gain can not be increased even when the channel between $S$ and $R$ is non-fading, which is because energy-exhausted phenomenon at $R$ becomes the bottleneck of whole system. Due to the limit of space, the demonstration is ignored here.

\emph{Corollary 2: Through cooperative protocol will occupy more time slot or frequency band, compared with traditional direct transmission protocol, there is a multiplicative performance gain in terms of outage probability, the gain of which is proportional to the energy-exhausted probability $p_{ex}$.}
\begin{proof}
Comparing the results in (\ref{eqn:outage probability in direct transmission}) with the first term in the right hand side of (\ref{eqn:outage probability evaluation}) can get above conclusion.
\end{proof}

Since energy-exhausted probability can be decreased to very small level in a well-designed energy harvesting system with the help of energy storage device \cite{Omur_11}, it is feasible and profitable to employ energy harvesting relay to improve system outage performance.

\section{Simulation Results}

\begin{figure}[!t]
\centering
\includegraphics[width=3.4 in]{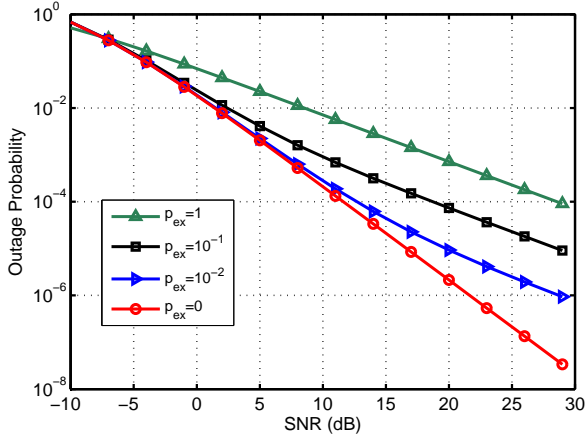}
\caption{The outage probability of cooperative networks as a function of SNR (namely $SNR=P_0/\sigma_0^2$) when the energy-exhausted probability $p_{ex}$ at $R$ is $1$, $10^{-1}$, $10^{-2}$ and $0$, respectively.}
\label{fig:Simulation_Result 1}
\end{figure}

It is assumed that the available frequency bandwidth for whole networks shown in Fig. \ref{fig:system framework} is $W=2$ MHz, and the minimum acceptable rate $R_0$ is $200$ kbps. With the assumption $P_s=P_r=P_0$, Fig. \ref{fig:Simulation_Result 1} illustrates the outage probability of cooperative networks as a function of SNR (namely $SNR=P_0/\sigma_0^2$) when the energy-exhausted probability $p_{ex}$ at $R$ is $1$, $10^{-1}$, $10^{-2}$ and $0$, respectively. Obviously, when $p_{ex}=1$, it is equivalent to the traditional system with direct transmission protocol. And when $p_{ex}=0$, it is equivalent to the system powered by constant power source which has been discussed in \cite{Laneman_2}. It can be observed from Fig. \ref{fig:Simulation_Result 1} that energy harvesting relay node can improve the outage performance of system significantly compared with traditional direct transmission protocol during the concerned SNR range. For example, outage probability can be decreased by an order of magnitude when the energy-exhausted probability is $p_{ex}=0.1$. The work in \cite{Omur_11} has indicated that $p_{ex}$ can be very small in a well-designed energy harvesting system. Thus, it is profitable to apply energy harvested relay node into wireless network to improve the system performance.

\begin{figure}[!t]
\centering
\includegraphics[width=3.4 in]{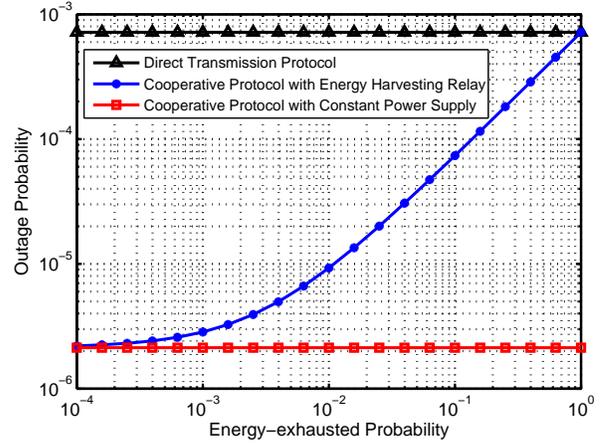}
\caption{The outage probability of cooperative system as a function of energy-exhausted probability at $R$ when the SNR is $20$ dB.}
\label{fig:Simulation_Result 2}
\end{figure}

Fig. \ref{fig:Simulation_Result 2} illustrates the outage probability of cooperative system as a function of energy-exhausted probability at $R$ when the SNR is $20$ dB. Besides, served as two comparisons, the outage probabilities of traditional direct transmission protocol and cooperative networks with constant power source are also given, which can be regarded as two extreme cases of cooperative networks with energy harvesting relay. It can be observed that cooperative network with energy harvesting bridges the two extreme cases. The system performance is improved apparently as the decrement of $p_{ex}$, especially when $p_{ex}\in [10^{-3},1]$.

\section{Conclusion}

This paper mainly investigates the benefit from cooperative transmission aided by energy harvesting relay node in terms of outage behavior in slow fading scenario. An on-off model is used to describe the energy flow harvested from surrounding environment. The explicit closed-form of outage probability of cooperative system is obtained as a function of energy-exhausted probability at relay node. By theoretical proofing and numerical simulation, it can be concluded that energy harvesting relay-aided cooperative transmission cannot increase the diversity gain compared with direct transmission protocol excepting the extreme case that energy-exhausted probability is zero. But multiplicative gain can be obtained in terms of outage performance in the concerned SNR range, the gain of which is proportional to the value of energy-exhausted probability. For instance, the outage probability can be decreased by an order of magnitude when energy-exhausted probability is $0.1$. Considering the fact that energy-exhausted probability in well-designed energy harvesting system is usually very small, it is potential and profitable to employ this technique to improve system outage performance in slow fading channel.

%

\ifCLASSOPTIONcaptionsoff
  \newpage
\fi

\end{document}